\documentclass[preprintnumbers,amsmath,amssymb,twocolumn,floatfix,prl]{revtex4}

\usepackage{dcolumn}
\usepackage{bm}
\usepackage{graphicx}

\begin{document}     
\title {Two-dimensional inversion symmetry as the fundamental symmetry
  of incompressible  quantum Hall fluids}
\author{ F. D. M. Haldane}
\affiliation{Department of Physics, Princeton University,
Princeton NJ 08544-0708}
\date{February 21, 2023}
\begin{abstract}
Two dimensional inversion symmetry ($180^{\circ}$ rotations in the
``Hall plane'' that hosts the incompressible electron fluid that exhibits
the quantized Hall effect) is identified as its fundamental unbroken
symmetry.  A consequence is that the integers $p$ and $q$ which define
both the Landau level filling factor $\nu$ = $p/q$ and the elementary
fractional charge $\pm e/q$ of topological excitations, cannot have a
common divisor greater than 2.

\end{abstract}
\maketitle
In the composite-boson picture\cite{girvinbook,gm,zhk,readboson} of the quantum Hall effect, $q$ London units
of magnetic flux (or $q$ independent one-electron orbitals per Landau
level) are ``attached'' to $p$ charge-$e$ particles to form
a fundamental unit (the ``composite boson'') which behaves like a
neutral particle that does not experience a Lorentz force as it moves;
pairs of these objects obey bosonic exchange statistics, allowing a
Bose condensation.    Unlike a standard Bose condensate, this
condensate is gapped (as a consequence of the Anderson-Higgs-like effect
of an emergent Abelian
Chern-Simons gauge field) and exhibits a quantized Hall conductance
$\sigma_H$ = $e^2\nu/2\pi \hbar$, with $\nu$ = $p/q$.

Initially it was thought that $p$ and $q$ were always coprime
(\textit{i.e.},
with greatest common divisor $\text{gcd}(p,q)$ equal to 1), leading to a so-called
``odd-denominator rule'' that asserted that $q$ must be odd for
fermionic FQH states.   The discovery of an ``even denominator
state'' at $\nu$ = 5/2 indicated that $p$ and $q$ are two independent
parameters that are not necessarily coprime, and indeed $q$ can be
independently determined by the elementary unit  $e/q$ of the charge of 
topological excitations of the fluid, which is $e/4$ in the case of
the  $(p,q)$ = $(2,4)$ Moore-Read ``Pfaffian'' state\cite{MR}.   A Pfaffian
state of the spin-polarized second Landau level (or its particle-hole
conjugate, the ``anti-Pfaffian''), plus a fully filled (both spin
components) lowest Landau level, are common explanations for the
observed $\nu$ = $5/2$ state.

Since this shows that  ${\rm gcd}(p,q)$ can be greater than one,
it is important to inquire if there are any  constraints on the common
divisor.  The
answer will be supplied by symmetry considerations.  While the
standard ``toy models'' of the FQHE (\textit{e.g.,}
Laughlin\cite{laughlin}, Moore-Read\cite{MR} wavefunctions)
incorporate a continuous $SO(2)$ rotation symmetry
around the normal to the quantum Hall surface, this is not a
protective symmetry (it can be removed by tilting the magnetic field away
from the normal to the surface, without harming the FQH effect).  Such a symmetry has no fundamental place in
condensed matter theories of electrons moving inside a fixed crystal
lattice, so it can be rejected as an unphysical (but useful) ``toy-model'' feature
that can  be  invoked to simplify calculations.

While $SO(2)$ symmetry can be dismissed as a non-generic feature, the
$C_2$ symmetry (two-fold, or $180^{\circ}$, rotation in the plane), is a
commonly-ignored symmetry which here is argued to be \textit{the}
fundamental underlying symmetry of  QHE states.     This is the 2D analog
of 3D inversion,  $\bm x \rightarrow -\bm x$, and does not require
specifying a rotation axis normal to the plane (it is also a 3D inversion
about a point on the plane, followed by mirror reflection in the plane). 
As part of the classification of QHE states,
the composite boson carries a parity quantum number for 2D
inversion about its center.

The ``clean limit'' of all uniform quantum Hall states has unbroken
translation and 2D inversion symmetry, and (when ``compactified'' on
the torus)  all members of the
topologically-degenerate multiplet of FQHE ground states 
have the same parity quantum number.   There is an energy gap separating
Bloch-type non-zero momentum eigenstates
which are not parity eigenstates from the ground-state
topological multiplet. Unbroken
inversion symmetry of uniform quantum Hall fluids constrains
$\text{gcd}(p,q)$ to be no greater than 2. To see this, consider the
many-body translational quantum numbers of a translationally-invariant
system on the torus\cite{haldanetorus}

The position $\bm x$ of a charge-$e$ particle on the Hall surface can
be written as the  sum of a ``guiding-center'' $\bm R$ plus the radius
vector  $\tilde{\bm R}$ of a closed Landau orbit around the guiding center.
The guiding-center coordinates $\bm R$ = $R^a\bm e_a$, with
contravariant
indices
$a$ $\in \{1,2\}$, obey the Heisenberg algebra
\begin{equation}
  [R^a,R^b] = -i\ell^2 \epsilon^{ab}
  \label{fr}
\end{equation}
where $2\pi \ell^2 $ = $2\pi  \hbar /eB > 0$ is the area of the plane
through which one London quantum of magnetic flux passes.
Here $\bm e_a$ are orthonormal tangent vectors where $\bm e_a \cdot
\bm e_b$ = $\delta_{ab}$, the Euclidean metric used to set up a 2D
Cartesian coordinate system, and $\epsilon^{ab}$ is an antisymmetric 2D Levi-Civita
symbol with a handedness (orientation) chosen to make $eB$ positive, where $F_{ab}$
= $\epsilon_{ab}B$ is the Faraday tensor in the 2D plane.
The dynamical momentum $p_a-eA_a$ is $\bar p_a^{(n)} + eF_{ab}\tilde R^b$,
 where $\bar{\bm p}^{(n)}$ is $\hbar$ times the time-averaged Bloch vector of
 the closed  semiclassical Landau orbit with index $n$ in the
 Brillouin zone of the underlying band structure.

The unitary one-particle translation operator $t(\bm a)$ is
$\exp (i\bm a \times \bm R/\ell^2)$, so $t(\bm a)\bm R$ =
$(\bm R + \bm a) t(\bm a)$, 
where $\bm a \times \bm R$ $\equiv$ $\epsilon_{ab}a^aR^b$,
and $\epsilon^{ac}\epsilon_{bd}$ = $\delta^a_b\delta^c_d$ $-$ $\delta^a_d\delta^c_b$.
The Hamiltonian is invariant under $\bm R_i \mapsto \bm a \pm  \bm R_i$,
all $i$, where $i$ is a particle label, \textit{i.e.}, translation and
2D inversion symmetry, which preserve the fundamental commutation
relations (\ref{fr}).

To treat a finite number of particles without
boundary effects, periodic boundary conditions can be applied to
``compactify'' the system on a 2D torus.
  $N$ particles will be placed on the torus with the boundary condition (pbc)
\begin{equation}
  t_i(\bm L_{mn}) |\Psi\rangle = (\eta_{m,n})^{N_{\Phi}}|\Psi\rangle
  \label{pbc}
\end{equation}
where $\bm L_{mn}$ = $m\bm L_1 +  n\bm L_2$, and $\eta_{m,n}$ = 1 if $m$ and $n$ are both even, and $-1$
otherwise, with
\begin{equation}
  \bm L_1 \times \bm L_2 = 2\pi N_{\Phi}\ell^2,
\end{equation}
where $N_{\Phi}$ is a positive integer.
There are $(N_{\Phi})^2$ independent unitary one-particle operators compatible
with the pbc, given by $t(L_{m,n}/N_{\Phi})$, with $1 \le m,n \le N_{\Phi}$.
It is convenient to represent these as $u(\bm k)$ =$\exp i \bm k\cdot
\bm R$, where $\bm k$ is a reciprocal vector satisfying $\exp (i\bm
k\cdot \bm L_{m,n})$ = 1 for all $\bm L_{m,n}$, where $\bm k$ $\in$
$\{\bm k_{m,n}\}$ = $\{m\bm G_1 + n\bm G_2\}$.
Any set of  $(N_{\Phi})^2$  reciprocal vectors $\{\bm k_i\}$
where, for $i \ne j$,  $\bm k_i - \bm k_j$ $\not \in$ $\{ N_{\Phi}\bm k\}$,  can be chosen as
a  (single-particle) ``Brillouin zone'' (BZ) that provides a complete
linearly-independent
set $\{ u(\bm k_i)\}$.

The number $N$ can be written
as $\sum_n N_n$, where $N_n$ are the number of particles in the
Landau-level with index $n$.  ``Landau-level index'' includes spin,
valley, layer labels, \text{etc}., {\textit i.e.}, anything other than
the guiding-center  that characterizes an electron state, so
\begin{equation}
  P_ne^{i\bm q\cdot \bm x}P_n = f_n(\bm q) P_ne^{i\bm q\cdot \bm R},
\end{equation}
where $f_n(\bm q)$ is a Landau-level form-factor, and $P_n$ is a
projection into Landau level $n$.

Let $\bar N_0$ = $gcd(N_{\Phi},
\{N_p\})$, so $N_{\Phi}$ = $q^0\bar N_0$ and $N_n$ = $p^0_n\bar N_0$,
with $p^0$ = $\sum_n p^0_n$.
The
labels $0$ are  used here  because this is a provisional reduction,
and  these
quantities may later be rescaled by an integer factor $r$:
\begin{equation}
\bar N_0 \mapsto \bar N = N_0/r,  \quad  q^0 \mapsto q = rq^0,
p_n^0 \mapsto p_n = rp_0.
\end{equation}
While the number $\bar N_0$ is mathematically motivated, the number $\bar N$
will be  physically motivated as the number of elementary units of the
FQH state
(``\textit{composite bosons}'', characterized by the integers  $\{q,
\{p_n\}\}$), which are analogous to the unit cells of a crystalline solid.
Continuity as a function of
system size will be achieved by choosing $\bar N$ to be integer, and
increasing it in unit steps $\bar N$ $\mapsto$ $\bar N$ + 1, 

The many-body eigenstates of the $N$-particle Hamiltonian can be
characterized\cite{haldanetorus} by a translational quantum number $\bm
K$ $\in$ $\{\bm k\}$.
 The quantities
\begin{equation}
  \{ T(\tfrac{L_{m,n}}{\bar N_0})\}, \quad T(\bm a) \equiv {\textstyle\prod_i} t_i(\bm a),
  \end{equation} are a mutually commuting set of unitary operators that can be
  simultaneously diagonalized:
  \begin{equation}
    T(\tfrac{\bm L_{m,n}}{\bar N_0})|\Psi_{n, \mu}(\bm K)\rangle
    = (\eta_{m,n})^{p^0q^0} e^{iq_0\bm K \cdot \bm L_{mn}}
    |\Psi_{n,\mu}\rangle;
  \end{equation}
  here $\mu$ labels a $q^0$-fold degeneracy, which can be resolved
  by choosing a primitive $\bm L_0$ = $\bm L_{m,n}$, with
  $\text{gcd}(m,n)$ = 1, so, for $\mu$ = $1,\ldots, q^0$, and $\phi$ = $2\pi/q^0$,
  \begin{equation}
    T(\tfrac{\bm L_0}{ N_{\Phi}})|\Psi_{n,\mu}(\bm K)\rangle = e^{i\mu \phi}
    (-1)^{p^0} e^{i\bm K\cdot \bm L_0}|\Psi_{n,\mu}(\bm
    K)\rangle. 
    \end{equation}
 
 The allowed values of $\bm K$ are reciprocal vectors
that take $(\bar N_0)^2$ distinct values in a ``many-particle BZ''  where $K$ is defined modulo $\{ \bar N_0\bm k\}$.
In the thermodynamic limit $\bar N_0\rightarrow \infty$, the
BZ expands to cover all reciprocal space, the spacing
of the lattice of allowed $\bm K$'s shrinks to become a continuous cover. 
The existence of a BZ is due to the property that, on the finite torus, the
``Bloch states'' are only invariant under discrete translations $\bm
L_{m,n}/\bar N_0$ that mutually commute and commute with the boundary conditions (\ref{pbc}).

The algebra of the one-electron operators $u_i(\bm k_{m,n})$ 
is an example of the general rank-$q^2$ Heisenberg algebra
\begin{align}
  u_{m,n} u_{m',n'} &= e^{i\pi  (p/q) (mn'-nm')} u_{m+m',n+n'} ,\nonumber
  \\
  (u_{m,n})^q &= (\eta_{m,n})^{pq} \openone,
  \end{align}
  where $u_{m,n}$ is unitary, $u_{0,0}$ = $\openone$ is the identity,
  and  $p,q$ are integers with $\text{gcd}(p,q)$ = 1.
  The unitary inversion operator $I$, with action
  $ Iu_{m,n}$ = $u_{-m,-n} I$ is given by
  \begin{equation}
     I = \tfrac{1}{q}\sideset{}{'}\sum_{m,n}I^{p,q}_{m,n} u_{m,n}
  \end{equation}
  where the sum is over \textit{any} choice of a linearly-independent set
  (BZ)  of $q^2$
  operators $u_{m,n}$ (such as the set  with $1\le m,n \le q)$), and
  is invariant under changes of BZ choice:
\begin{equation}
I^{p,q}_{m,n} = \tfrac{1}{2}(1 - (-1)^{p+q}) +
\tfrac{1}{2}(1 - (-1)^p)\eta_{m,n}.
\label{invert}
\end{equation}

The  one-electron inversion operator $I_i$ that acts only on particle
$i$ is constructed using (\ref{invert}) with $p$ = 1, $q$ = $N_{\Phi}$,
and the many-body inversion operator $I$ is $\prod_iI_i$, with the action
\begin{equation}
  I|\Psi_{n,\mu}(\bm K)\rangle = |\Psi_{n,-\mu}(-\bm K)\rangle.
\end{equation}
A ``center of mass'' inversion operator can be constructed as
\begin{equation}
  I_{\text{cm}}= \frac{1}{N_0N_{\Phi}}\sideset{}{'}\sum_{\bm k}
  I_{\text{cm}}(\bm k) U(\bm k), \quad U(\bm k) \equiv {\textstyle\prod_i} u_i(\bm k),
\end{equation}
where $I_{\text{cm}}(\bm k_{m,n})$ = $I^{p^0,q^0}_{m,n}$,
and $I_{\text{cm}}U(\bm k)$ = $U(-\bm k)I_{\text{cm}}$; the
primed sum is over the one-electron BZ. 

The condition for a Bloch eigenstate to also be an inversion
eigenstate is that, for all $\bm L_{m,n}$,
\begin{equation}
  e^{iq^0\bm K\cdot \bm L_{mn}} = (\sigma_1)^m(\sigma_2)^n, \quad (\sigma_i)^2 =
(\sigma_i)^{\bar N_0} = 1.
\end{equation}
For odd $\bar N_0$, only the case $(\sigma_1,\sigma_2)$ = $(1,1)$ is
possible, so $\bm K$ = $0\mod \bar N_0\bm k$.    For even $\bar N_0$
there are  three
other possibilities $(-1,1)$, $(1,-1)$ and $(-1,-1)$. 
For these special values of $\bm K$, a many-body inversion quantum
number  $\xi$ = $\pm 1$ is given by
\begin{equation}
I|\Psi_{n,\mu}(\sigma_1,\sigma_2)\rangle = \xi_n
I_{\text{cm}}|\Psi_{n,\mu}(\sigma_1,\sigma_2)\rangle .
\end{equation}

A ``topological multiplet'' of ground states on the
translationally-invariant
torus is a group
of  inversion-symmetric
states that, while not generally  exactly degenerate, have energy
splittings that are exponentially small at large system size, and are
separated from other energy levels
by a spectral gap that remains finite in the thermodynamic limit.
All members of the multiplet have an exact $q^0$-fold
``center of mass''
degeneracy,  and for large $\bar N_0$,
tiny energy splittings between members of a generic
topological multiplet remain tiny  when the system is
perturbed;
this distinguishes a true topological multiplet from the case
where two such multiplets are quasi-degenerate, either for symmetry
reasons or accidentally, as they will split apart when perturbed.

Any linear combination of the
states in a topological multiplet should be a representative state, so  all
members of the multiplet must have the same inversion quantum number
$I_n$.  If some members of the multiplet occur at $\bm K$ $\ne
\bm 0$, the number of states $d_1$ at each point $(1,-1)$, $(-1,1)$, $(1,1)$
must be equal,  so the topological degeneracy has the form
$d$ = $q_0(d_0 + 3d_1)$; for example, for the Laughlin\cite{laughlin} ($n$=1),
Moore-Read\cite{MR} ($n$ = 2),
Read-Rezayi\cite{RR} ($n =3, 4 ,\ldots$) sequence  (which for $n > 1$ are
non-Abelian states with a neutral
$Z_n$ parafermion quasiparticle), $(d_0,d_1)$ = $(\tfrac{1}{2}(n+1),0)$
for odd $n$, but is given  by $(r, r+1)$ for $n$ = $4r + 2$, and by
 $(r+1, r)$ for $n$ = $4r$.

Since one should be considering
an arbitrary state in the multiplet, it is inappropriate to split up the
multiplet by the values of $(s_1,s_2)$.   The existence of a non-zero
$d_1$, which can only occur when $\bar N_0$ is even,
indicates that $\bar N_0$  is not the correct value of $\bar N$, now
interpreted as the number of elementary FQH units (``composite
bosons'').
Reducing $\bar N_0$ by a factor of $\frac{1}{2}$ to $\bar
N$ = $\tfrac{1}{2}\bar N_0$ and doubling $q$ and $p_n$ to
$ 2q^0$ and $2p^0_n$, maps the ``many-particle BZ'' into four copies
of the quarter-BZ, and the inversion-symmetric
states now all obey
\begin{equation}
  T(\tfrac{\bm L_{m,n}}{\bar N}) |\Psi_{n,\mu}(\bm 0)\rangle = (\eta_{m,n})^{pq}|\Psi_{n,\mu}(\bm
0)\rangle,
\end{equation}
as in the case $\bar N$ = $\bar N_0$.
The presentation of the excitation spectrum in the quarter BZ centered
on the now-unique inversion-symmetric point now also provides a
meaningful spectrum as a function of momentum when $d_1$ $\ne$ 0, using
the  Wigner-Seitz construction of the BZ with the
Euclidean metric $\delta_{ab}$,   where $\bm k$ is chosen so
$|\bm k|^2$ = $\min_{\bm k'} \delta^{ab}(k_a- \bar Nk'_a)(k_b-\bar Nk'_b)$.  

There are just two possibilities, whether or not the FQH multiplets
have non-zero $d_1$, so there just two possibilities for
$\text{gcd}(p,q)$: 1 or 2.   Thus the fundamental property that the
FQH topological multiplet must have a definite (unbroken) symmetry (parity) under
2D inversion ($180^{\circ}$ rotation in the plane), provides a
powerful constraint on the possible values of $p$ and $q$, given $\nu$
= $p/q$.   The condition that (clean-limit) incompressible quantum Hall states have
unbroken 2D inversion symmetry directly relates to their property
that there is  a bulk gap for excitations carrying electric polarization tangent
to the ``Hall surface'' that exhibits the quantized Hall effect, showing
that this symmetry is fundamental to the quantum Hall fluids. 

Much work on the FQH involves model systems that enlarge 2D inversion symmetry into a continuous (planar, or azimuthal)
$SO(2)$ rotation symmetry of the model Hamiltonian, which in particular
is a feature of  important ``toy model'' wavefunctions (like the Laughlin state)
which are constructed using ``conformal blocks" from (Euclidean)
conformal field theory (cft), which is based on a further enlargement of
this symmetry.   Since continuous rotation symmetries are  here
asserted to be unphysical
ingredients of FQH theory, one may ask why cft has been so useful.
First, none of its useful ingredients turn out to rely on its
conformal symmetry: it is its \textit{non-conformal} topological aspects
(braiding relations) that  are generic.   Second, the incompressible
FQH fluid does not transmit force though its bulk or support pressure,
so its stress tensor is traceless, $\sigma^a_a$ = 0, which is the
Euclidean version of the tracelessness of the stress-energy tensor in
cft.  Finally, the \textit{momentum} (but \textit{not} energy) density  on the edge
of FQH liquids obeys the (chiral) Virasoro algebra, generally with a
(signed) central charge, providing a further connection to cft.

The generator $L$ = $-B\delta_{ab}q^{ab}$ of $SO(2)$ rotations in
these models is defined using
the Euclidean metric and
\begin{equation}
 \quad q^{ab} = \tfrac{1}{2}e\sum_i
  \tfrac{1}{2}\{\tilde R_i^a,\tilde R_i^b\} -  \tfrac{1}{2}\{R_i^a,R^b_i\},
 \end{equation}
 where $q^{ab}$ is the total (``primitive'', not ``traceless'')
 \textit{electric quadrupole moment},
 and $\{A,B\}$ is the anticommutator $AB + BA$.
 The expression for the total angular momentum is a sum of two independent
 terms, respectively involving the Landau orbit vectors $\{\tilde {\bm R}_i\}$
  and the guiding
 centers $\{\bm R_i\}$.

 In terms of occupations $n_{m,n}$ of
 electron orbitals labeled by ${m,n}$, with  0 $\le m,n$, and
 parities $(-1)^n(-1)^m$, where $n$ is the index of Newtonian  Landau levels
 $\hbar \omega_c (n + \tfrac{1}{2})$,
  $L$ = $\hbar\sum_{m,n} (m-n)n_{m,n}$.
 For a finite circular droplet of $\bar N$ composite
 bosons, with unexcited edge states, the angular momentum has the form
 \begin{equation}
   L = \hbar (\tfrac{1}{2}k\bar N^2 + S\bar N), \quad (-1)^{2S} =
   (-1)^{k},
 \end{equation}
 where $k$ = $pq$ is the integer index of the emergent $U(1)_k$
 Chern-Simons
 gauge field that describes the Berry phase accumulated during the
 motion of one composite boson through the background of the others,
 and $\hbar S$ is the \textit{intrinsic  orbital angular momentum of the composite
 boson} around its inversion center, which can be written as
 \begin{equation}
   S = S_{\text{GC}} + \sum_n s_np_n
\end{equation}
Here $\hbar s_n$ are the Landau orbit eigenvalues of
$-\frac{1}{2}eB\delta_{ab}\tilde R^a\tilde R^b$, where (for electrons
with non-relativistic Newtonian dynamics)  $-s_n$ =
$n+\tfrac{1}{2}$, (with parity $\xi_n$ = $(-1)^n$), labeled by $n$ = $0,1,2\ldots$, and, for $\bar N$ = 1,
\begin{equation}
  S_{\text{GC}} = \sum_{m=0}^{q-1}(m +\tfrac{1}{2})(n_m - p/q), \quad n_m = \sum_n
  n_{m,n},
  \end{equation}
is the \textit{``guiding-center spin''}, which characterizes the
deviation of the electron distribution inside the composite boson from
uniform occupation numbers (\textit{e.g.} in the Laughlin states, $n_m$ =
$\delta_{m,0}$).
An important property of $S_{\text{GC}}$ is that it has no contributions from
filled (or empty) Landau levels, and that it is odd under
particle-hole transformations of partially-filled Landau levels.

Note that the composite boson orbital  angular momentum $S$, and its
components $S_{\text {GC}}$ and $s_n$, all take standard integer or
 half-integer values in units of $\hbar$.
 It is related to previously-defined quantities by $S$ = $p\bar s$ =
 $-\tfrac{1}{2}p\mathcal S$, where $\bar s$ was defined by Read\cite{readvisc}
 as the ``intrinsic orbital angular momentum per electron''
 in his study of Hall viscosity, and
 $\mathcal S$ is the ``shift'' introduced by Wen and Zee\cite{wenzee} in connection
 with models of the QH fluid on the surface of a sphere\cite{haldanesphere}. 
 Both these constructs are generically given by arbitrary fractions,
 which is a clue that they are not in fact the physically meaningful
 quantity: it is not the electron itself that is the fundamental unit, it is the \textit{composite boson},
 which contains $p$ electrons that is fundamental.

 The condition that the composite boson has bosonic
 exchange statistics is
 \begin{equation}
   (-1)^k = (\zeta)^p = (-1)^{2S}, \quad \zeta = -1,
\label{ed}
 \end{equation}
 where the exchange factor $(-1)^k$ generated by the $U(1)_k$
 Chern-Simons field cancels
 the exchange factor for exchange or two groups of $p$ fermions.
(If the charge-$e$ particles were bosons, as in $p$ = 1 Laughlin
states with even $q$, $\zeta$ would be $+1$.)  The composite boson 
therefore has integral or half-integral S, depending on whether the
fermion parity of  the  particles it contains is even or odd.   Since
the Landau orbits have half-integral angular momentum, this
in turn means that $S_{\text {GC}}$ is always integral when the charge-$e$
particles are fermions,  and would be  integral or half-integral in boson systems,
depending on whether the number of charge-$e$ bosons is even  or odd.
The selection rule (\ref{ed}) makes $S_{\text {GC}}$ an integer for
fermion systems.

If $\text{gcd}(p,q)$ = 1,
($(p,q)$ = $(p_0,q_0)$)
the rule (\ref{ed}) for fermions rules out an even $q$ = $q_0$, so any
``even denominator'' FQH states with $\nu$ =  (coprime) $p_0/q_0$ with even $q_0$
must have $\text{gcd}(p,q)$ = 2.   A $\nu$ = $5/2$  FQH state,
where the $n$ = 0 Landau level (both spin components) is fully filled,
can be viewed as a direct product of three decoupled FQH fluids:
$U(1)_1\otimes U(1)_1\otimes U(1)_8$, where the $n=1$ level hosts
a $(p,q)$ = $(2,4)$ fluid with charge $\pm e/4$ elementary
topological excitations.

The composite boson carries electric charge $e^*$ = $pe$.
All the fundamental QH properties can be expressed in
terms of the basic composite-boson parameters $e^*$, $k$, and (in
models with $SO(2)$ symmetry)
$S$ \textit{without referencing the parameters $p$ and $q$
which characterize its microscopic structure in terms of electrons}.
For example, the
Hall conductance is $\sigma_H$ = $(e^*)^2/2\pi \hbar  k$, and the
topological excitations carry charge in units $e^*/k$.  The
$SO(2)$-symmetric Hall
viscosity,  obtained  by Read\cite{readvisc} can be written as $\eta_H$ =  $e^*BS/4\pi
\hbar k$.

While the integer-valued orbital angular momentum $S_{\text{cb}}$ of the composite boson is a
  toy-model feature, its \textit{intrinsic parity} $(-1)^{S+\tfrac{1}{2}k}$ =
  $\xi_{\text {GC}}\prod_n(\xi_n)^{p_n}$, with $\xi_{\rm GC}$ = 
  $(-1)^{S_{\text{GC}} + \tfrac{1}{2}(k-p)}$ is a
  generic feature.
    The parity quantum number of the ground state
  topological-multiplet  of $\bar N$ composite bosons on the torus is
  then
  given by
  \begin{equation}
    I_{\text{cm}}I =(-1)^{\tfrac{1}{2}k\bar N(\bar N-1)}
      (-1)^{(S+\tfrac{1}{2}k) \bar N},
    \end{equation}
and the natural sequence for  approaching the thermodynamic limit is
the sequence of consecutive integers $\bar N$.
(It follows from the definitions given earlier that  $S +
\tfrac{1}{2}k$  is an integer.)

In summary, 2D inversion symmetry
($180^{\circ}$ rotation in the Hall plane) is argued to be the fundamental unbroken
symmetry of 2D quantum Hall fluids, unlike a commonly invoked continuous
$SO(2)$ symmetry which is merely a convenient but unphysical
feature of  simplified ``toy models''.   The 2D inversion
symmetry limits the greatest common divisor of the two
numbers $p$ and $q$  to no more than 2: these numbers
define the electron filling factor $\nu$ =
$p/q$, the Chern-Simons index $k$= $pq$, the charge $e^*$ = $pe$
of the elementary unit (composite boson) of the fluid and the
elementary fractional charge $e^*/k$ = $e/q$.   The so-called ``even
denominator''  FQH states are fermionic systems with $\text{gcd}(p,q)$
= 2.

This research was supported, in part, by the US Department of Energy,
Basic Energy Sciences grant DE-SC0002140.
It was brought to
conclusion, after the end
of that support, with partial support by NSF through the Princeton
University (PCCM) Materials Research Science and Engineering Center
DMR-2011750.

\end{document}